\documentclass[10pt, conference, letterpaper]{ IEEEtran}

\usepackage{cite}
\usepackage{amsmath,amssymb,amsfonts}
\usepackage{amsthm}
\usepackage{graphicx}
\usepackage{textcomp}
\usepackage{algorithm}
\usepackage{algorithmic}
\usepackage{xcolor}
\usepackage{booktabs}
\usepackage{tikz}
\usepackage{comment}
\usetikzlibrary{arrows.meta,positioning}
\usepackage{array}
\usepackage{graphicx} 
\usepackage{subcaption}
\usepackage[bookmarks=false]{hyperref}
\theoremstyle{definition}

\def\BibTeX{{\rm B\kern-.05em{\sc i\kern-.025em b}\kern-.08em
    T\kern-.1667em\lower.7ex\hbox{E}\kern-.125emX}}

\begin{document}
\title{FedMaJ: Ordinal Meta-Aggregation for Federated Learning Using Majority-Judgment Voting}

\author{Driss Choukri$^{\bullet\circ}$,
Essaid Sabir$^{\bullet}$,
Halima Elbiaze$^{\diamond}$
and Abdelkrim Haqiq$^{\circ}$\\
$^{\bullet}$Department of Science and Technology,
TÉLUQ, University of Quebec, Montreal, H2S 3L4, Canada\\
$^{\circ}$Computer Networks, Mobility and Modeling Laboratory,
FST, Hassan I University of Settat, Morocco\\
$^{\diamond}$Department of Computer Science,
University of Quebec at Montreal, Montreal, Canada\\
Emails: driss.choukri.doc@uhp.ac.ma, essaid.sabir@teluq.ca,\\
elbiaze.halima@uqam.ca, abdelkrim.haqiq@uhp.ac.ma}

\maketitle

\begin{abstract}
Federated learning traditionally relies on a fixed aggregation rule despite shifting client state and wireless dynamics. We present FedMaJ, a framework that dynamically selects aggregation rules via Majority Judgment voting. During voting rounds, clients evaluate candidate models across performance, resource utilization, channel condition, and fairness metrics. The server adopts the majority-preferred rule and retains it until the next vote, scheduled dynamically based on accuracy trends. Evaluated in over-the-air federated learning with explicit energy and resource cost modeling, synthetic and Fashion-MNIST experiments show that FedMaJ achieves competitive learning performance and superior aggregation-influence fairness. Under extreme-evaluation perturbations, Majority Judgment yields significantly more stable rule selections than mean-based scoring, establishing ordinal voting as a robust control layer for adaptive aggregation.
\end{abstract}


\section{Introduction}

Federated learning (FL) enables distributed clients to train a shared model without transmitting raw data. Its performance depends on how client updates are aggregated. Existing methods account for dataset size~\cite{McMahan2017}, data quality~\cite{zhang2025dual}, resource availability~\cite{nishio2018client}, wireless conditions~\cite{amiri2020federated}, or participation history~\cite{tan2022adafed}. However, the preferred criterion may change as learning progresses and client resources, channels, and participation patterns evolve. In over-the-air FL (OTA-FL), this choice also interacts with transmit scaling, receiver denoising, and aggregation distortion~\cite{yang2020ota_fl,cao2022air_fedavg_power_control}. Adaptive aggregation addresses part of this challenge. For example, FedMcon learns an aggregator through a meta-controller trained on proxy data~\cite{shen2025fedmcon}. Here, we investigate a complementary approach: selecting among a finite library of aggregation rules using client evaluations. Averaging numerical evaluations can be sensitive to extreme scores and requires comparable scoring scales. This motivates an ordinal decision mechanism that reflects broad client support.

We propose FedMaJ, a client-informed framework for dynamic aggregation-rule selection based on Majority Judgment~\cite{MJbook}. At voting rounds, selected clients digitally upload their updates, and the parameter server (PS) constructs and broadcasts one candidate model per rule. Clients evaluate these candidates locally and assign ordinal grades using learning, communication, resource, and fairness indicators. The PS selects the rule with the highest majority grade, resolves ties through iterative median comparison, and retains the rule until the next vote. An accuracy-dependent voting interval adjusts the frequency of candidate evaluation and selection.

FedMaJ separates the choice of aggregation policy from its execution. It supports conventional digital FL and is instantiated here in OTA-FL, where the selected logical weights interact with wireless transmission and aggregation distortion. By involving clients in policy selection, FedMaJ makes their evolving preferences part of the aggregation decision rather than relying solely on server-defined criteria. The main contributions are:
\begin{itemize}
    \item \textbf{Client-informed ordinal selection:}
    We formulate aggregation-rule selection as a repeated decision in which clients grade candidate models and Majority Judgment identifies the winning rule.
    
    \item \textbf{OTA-FL integration and overhead modeling:}
    We integrate rule selection with wireless-aware indicators and adaptive voting, and characterize its communication, computation, and energy costs.
    
    \item \textbf{Experimental evaluation:}
    Synthetic and Fashion-MNIST experiments compare FedMaJ with fixed rules, random selection, and average-score selection, assessing learning performance, influence fairness, system costs, client satisfaction, and sensitivity to extreme evaluations.
\end{itemize}
\section{Majority-Judgment-Based Federated Learning}

\subsection{General FL Setting and Candidate Rules}

Consider a PS and clients $\mathcal{K}=\{1,2,\ldots,K\}$. At communication round $t$, a subset $\mathcal{S}_t\subseteq\mathcal{K}$ participates. Client $k$ owns dataset $\mathcal{D}_k$ with size $n_k$ and computes a local update $\Delta_{t,k}$ from the current global model. FedMaJ assumes a finite candidate set $\mathcal{A}=\{A_1,A_2,\ldots,A_M\}$, and $M=|\mathcal A|$. Each rule $A_m$ produces logical aggregation weights $p_{t,k}^{(m)}$ satisfying $p_{t,k}^{(m)}\geq 0$ and $\sum_{k\in\mathcal{S}_t}p_{t,k}^{(m)}=1$. If rule $A_m$ is selected, the desired aggregate update is $\Delta_t^{(m)}=\sum_{k\in\mathcal{S}_t}p_{t,k}^{(m)}\Delta_{t,k}$ followed by the global update $\mathbf{w}_{t+1}=\mathbf{w}_t+\Delta_t^{(m)}$ or an equivalent model-averaging form. Table~\ref{tab:candidate_rules} summarizes the candidate rules used by FedMaJ and provides representative references for their underlying design principles. The cited works support these principles but do not necessarily use the exact proxy formulations adopted here.
Each rule defines a nonnegative score $\omega_{t,k}^{(m)}$,
normalized as $p_{t,k}^{(m)}=\omega_{t,k}^{(m)}/
\sum_{j\in\mathcal S_t}\omega_{t,j}^{(m)}$.

\begin{table}[!t]
\centering
\caption{Candidate rules and unnormalized scores.}
\label{tab:candidate_rules}
\footnotesize
\renewcommand{\arraystretch}{1.25}
\setlength{\tabcolsep}{3pt}
\begin{tabular}{@{}cll@{}}
\toprule
\textbf{Rule} & \textbf{Principle}
& \textbf{Score $\omega_{t,k}^{(m)}$} \\
\midrule
$A_1$ & Dataset size~\cite{McMahan2017}
& $n_k$ \\

$A_2$ & Uniform influence
& $1$ \\

$A_3$ & Loss awareness~\cite{zhang2025dual}
& $\ell_{t,k}+\epsilon_\ell$ \\

$A_4$ & Resources~\cite{nishio2018client,sun2022dynamic_scheduling_airfeel}
& $\frac{n_kB_{t,k}}{\widehat E_{t,k}+\epsilon_E}$ \\

$A_5$ & Link reliability~\cite{amiri2020federated}
& $n_kc_{t,k}$ \\

$A_6$ & OTA distortion~\cite{cao2022air_fedavg_power_control,azimi2024wafel}
& $\frac{n_k}{d_{t,k}+\epsilon_d}$ \\

$A_7$ & Participation~\cite{tan2022adafed,hamidi2025ota_ffl}
& $\frac{n_k}{N_{t,k}+\epsilon_N}$ \\
\bottomrule
\end{tabular}
\end{table}

Here, $\ell_{t,k}$ denotes local loss, $B_{t,k}$ residual
resources, $\widehat E_{t,k}$ estimated client cost,
$c_{t,k}$ link reliability, $d_{t,k}$ a distortion or
communication-cost indicator, and $N_{t,k}$ past participation
count. The $\epsilon$ terms are positive stabilizers.
All quantities are available before rule selection; in particular,
$\widehat E_{t,k}$ does not depend on the realized cost of the
rule selected in the same round.
Algorithm~\ref{alg:fedmaj} summarizes the procedure.

\subsection{Majority-Judgment Rule Selection}

At a voting round $t$, selected clients digitally upload their local updates through orthogonal links, allowing the PS to reuse each update under all $M$ candidate weight vectors. Dataset sizes, battery states, current channel indicators, and previous participation or influence are available before grading; post-round battery and influence updates are excluded. FedMaJ supports any finite ordered grade set $\mathcal{T}_n=\{T_1,T_2,\ldots,T_n\}$, with
$T_1 \prec T_2 \prec \cdots \prec T_n$,
where $T_1$ and $T_n$ are the lowest and highest grades. Each selected client assigns one grade to each candidate rule $A_m$, and Majority Judgment selects the winner using the grade distributions and their medians rather than arithmetic averages.

At a voting round, after receiving the individual updates, for any candidate $m \in \{1,2,\ldots,M\}$, the PS constructs
\begin{equation}
\Delta_t^{(m)}=\sum_{j\in\mathcal S_t}p_{t,j}^{(m)}\Delta\mathbf w_{t,j},\qquad
\mathbf w_{t+1}^{(m)}=\mathbf w_t+\Delta_t^{(m)}
\label{eq:candidate_global_model}
\end{equation}
The PS broadcasts all $M$ full candidate models to every selected client. Client $k$, not the PS, then evaluates each model on its private dataset $\ell_{t,k}^{(m)}=F_k\!\left(\mathbf w_{t+1}^{(m)}\right)
=\frac{1}{n_k}\sum_{\xi\in\mathcal D_k}f\!\left(\mathbf w_{t+1}^{(m)};\xi\right)$. Each candidate is evaluated locally on private data; raw datasets
are never transmitted.

For scalars $\{x_i\}_{i\in\mathcal I}$, let
$x_{\min}=\min_{j\in\mathcal I}x_j$ and
$x_{\max}=\max_{j\in\mathcal I}x_j$. Define
$\mathcal N_{\mathcal I}(x_i)=(x_i-x_{\min})/(x_{\max}-x_{\min})$
when $x_{\max}>x_{\min}$, and $0.5$ otherwise; reverse
normalization is $\mathcal N_{\mathcal I}^{-}(x_i)=1-\mathcal N_{\mathcal I}(x_i)$.
For each candidate, use
$\widetilde p_{t,k}^{(m)}=\mathcal N_{\mathcal S_t}(p_{t,k}^{(m)})$
and $\widetilde\ell_{t,k}^{(m)}=\mathcal N_{\mathcal S_t}(\ell_{t,k}^{(m)})$.
For $x\in\{c,B,I\}$, write
$\widetilde x_{t,k}^{-}=\mathcal N_{\mathcal S_t}^{-}(x_{t,k})$,
where $I_{t,k}$ is cumulative aggregation influence before round
$t$; participation count $N_{t,k}$ may serve as a proxy when
influence history is unavailable. The learning indicator
$Q_{t,k}^{(m)}=1-|\widetilde p_{t,k}^{(m)}-\widetilde\ell_{t,k}^{(m)}|$
matches influence to learning need, with larger loss indicating
greater need. Communication and resource indicators,
$C_{t,k}^{(m)}=1-\widetilde p_{t,k}^{(m)}\widetilde c_{t,k}^{-}$
and $R_{t,k}^{(m)}=1-\widetilde p_{t,k}^{(m)}\widetilde B_{t,k}^{-}$,
penalize high influence on unreliable links and resource-depleted
clients. The fairness indicator
$F_{t,k}^{(m)}=1-|\widetilde p_{t,k}^{(m)}-\widetilde I_{t,k}^{-}|$
favors previously underrepresented clients.
The combined grading score is
\begin{equation}
U_{t,k}^{(m)}
=\alpha_Q Q_{t,k}^{(m)}+\alpha_C C_{t,k}^{(m)}
+\alpha_R R_{t,k}^{(m)}+\alpha_F F_{t,k}^{(m)},
\label{eq:normalized_grade_score}
\end{equation}
where $\alpha_Q,\alpha_C,\alpha_R,\alpha_F\geq 0$ and
$\alpha_Q+\alpha_C+\alpha_R+\alpha_F=1$. Before grading, $U_{t,k}^{(m)}$ is clipped to $[0,1]$. For a grade set with $n$ grades, define the threshold vector $0=\theta_0 < \theta_1 < \cdots < \theta_{n-1} < \theta_n=1$. The grading map is $\Gamma_n(U)=T_i$ when
$\theta_{i-1}\leq U<\theta_i$, with $\Gamma_n(1)=T_n$.
Thus, each scalar score is mapped to an ordinal grade by $g_{t,k}^{(m)}=\Gamma_n\!\left(U_{t,k}^{(m)}\right)$. The PS collects $\mathcal{G}_{t}^{(m)}=\{g_{t,k}^{(m)}:k\in\mathcal{S}_t\}$ and selects
\begin{equation}
    A_t^\star=
    \arg\max_{A_m\in\mathcal{A}}\mathrm{MJ}\big(\mathcal{G}_{t}^{(m)}\big).
    \label{eq:selected_rule}
\end{equation}
Here, $\mathrm{MJ}$ ranks candidates by their median grade.
Ties are resolved by repeatedly removing one occurrence of the
common median from each tied candidate and comparing the
remaining medians.

\subsection{Client Satisfaction with the Selected Rule}
\label{subsec:client_satisfaction}

Let $\iota(T_i)=i$. For $M\geq2$, client $k$'s satisfaction
with any candidate $A$ is
\begin{equation}
S_{t,k}(A)=1-\frac{1}{M-1}\sum_{m=1}^{M}
\mathbb I\!\left[\iota(g_{t,k}^{(m)})>\iota(g_{t,k}^{(A)})\right].
\label{eq:client_satisfaction}
\end{equation}
It lies in $[0,1]$ and equals one when $A$ receives the client's
highest grade, including ties. This diagnostic does not affect selection.
At voting rounds, write $S_{t,k}=S_{t,k}(A_t^\star)$ and
$Z_v=\sum_{t=1}^{T}v_t|\mathcal S_t|$. Mean satisfaction and
top-choice rate are
\begin{align}
\overline S_{\mathrm{client}}
&=Z_v^{-1}\sum_{t=1}^{T}v_t\sum_{k\in\mathcal S_t}S_{t,k},
\label{eq:overall_client_satisfaction}\\
P_{\mathrm{top}}
&=Z_v^{-1}\sum_{t=1}^{T}v_t\sum_{k\in\mathcal S_t}
\mathbb I[S_{t,k}=1].
\label{eq:client_top_choice_rate}
\end{align}
For all-round comparisons, set
$S_{t,k}^{\mathrm{active}}=S_{t,k}(\widetilde A_t)$ for the
active rule $\widetilde A_t$, and average as
\begin{equation}
\overline S_{\mathrm{active}}
=\frac{\sum_{t=1}^{T}\sum_{k\in\mathcal S_t}S_{t,k}^{\mathrm{active}}}
{\sum_{t=1}^{T}|\mathcal S_t|}.
\label{eq:overall_active_rule_client_satisfaction}
\end{equation}
These all-round grades are reconstructed offline for fixed and
adaptive methods, adding no operational communication, energy,
latency, or voting overhead.

\subsection{Dynamic and Resilient Voting Mechanism}

To reduce decision-layer overhead, FedMaJ adapts the voting interval to
the accuracy margin:
\begin{equation}
\tau_t^{\mathrm{vote}}
=
\min\left\{
\tau_{\max},
1+
\left\lfloor
\frac{
[\mathrm{Acc}_{t-1}-\mathrm{Acc}_{\min}]^+
}{
\Delta_{\mathrm{acc}}
}
\right\rfloor
\right\},
\label{eq:adaptive_voting_interval}
\end{equation}
where $[x]^+=\max\{x,0\}$, $\mathrm{Acc}_{\min}$ is the required
accuracy, $\Delta_{\mathrm{acc}}>0$ is the accuracy-margin step, and
$\tau_{\max}$ is the maximum voting interval. Here,
$\mathrm{Acc}_{t-1}$ is the accuracy of the current global model
$\mathbf w_t$; for $t=1$, the PS evaluates the initial model to obtain
$\mathrm{Acc}_0$. This accuracy is supplied by the ordinary global-model
monitoring mechanism and is distinct from the candidate losses used in
client grading.

When $\mathrm{Acc}_{t-1}\leq\mathrm{Acc}_{\min}$, voting occurs every
round. Above this threshold, each additional $\Delta_{\mathrm{acc}}$
increases the interval by one, up to $\tau_{\max}$, thereby reducing
candidate-evaluation, reporting, broadcasting, and decision overhead.

Let $t_{\mathrm{last}}$ be the most recent voting round. With $v_1=1$,
the voting indicator for $t\geq2$ is
\begin{equation}
v_t
=
\begin{cases}
1, & t-t_{\mathrm{last}}\geq\tau_t^{\mathrm{vote}},\\
0, & \text{otherwise},
\end{cases}
\label{eq:adaptive_voting_indicator}
\end{equation}
and $t_{\mathrm{last}}\leftarrow t$ whenever $v_t=1$.

At a voting round, Majority Judgment selects
$\widetilde A_t=A_t^\star$ according to \eqref{eq:selected_rule};
otherwise, the PS retains
$\widetilde A_t=\widetilde A_{t-1}$. The resulting logical aggregate is
\begin{equation}
\Delta_t
=
\sum_{k\in\mathcal S_t}
p_{t,k}^{(\widetilde A_t)}\Delta_{t,k}.
\end{equation}
By relying on median-supported grades rather than average scores,
Majority Judgment limits the effect of extreme evaluations and favors
broadly supported rules. Clients are assumed to follow the prescribed
grading protocol; strategic manipulation and incentive-compatible
voting are left for future work.

\subsection{Conventional, Hybrid, and OTA-FL Interpretations}

In conventional digital FL, the PS directly applies the selected weights
$p_{t,k}^{(\widetilde{A}_t)}$ to the received updates. In wireless or
hybrid FL, the grades and indicators $c_{t,k}$ and $d_{t,k}$ can also
capture link reliability, delay, packet loss, quantization, and
retransmissions.

In OTA-FL, the selected logical weights are realized through
physical-layer variables. For a single-antenna system,
\begin{equation}
    \widehat{\Delta}_t
    =
    a_t
    \sum_{k\in\mathcal{S}_t}
    h_{t,k}b_{t,k}
    p_{t,k}^{(\widetilde{A}_t)}
    \Delta_{t,k}
    +
    a_tz_t,
    \label{eq:ota_logical_physical_model}
\end{equation}
where $h_{t,k}$, $b_{t,k}$, $a_t$, and $z_t$ denote the OTA channel,
transmit scaling, receiver denoising, and noise, respectively. Digital
grade reporting and control broadcasting use
$h_{t,k}^{\mathrm{ul}}$ and $h_{t,k}^{\mathrm{dl}}$. For multiple
antennas, $h_{t,k}$ is replaced by
$\mathbf{m}_t^H\mathbf{h}_{t,k}$.

Here, $c_{t,k}$ represents link reliability, while $d_{t,k}$ measures
the mismatch between desired and realized aggregation coefficients.
Thus, OTA-FL candidates jointly account for logical weights, wireless
controls, energy, and aggregation distortion, covering representative
OTA-FL designs
\cite{yang2020ota_fl,cao2022air_fedavg_power_control,azimi2024wafel}.
The remainder focuses on this OTA-FL setting.
\section{Proposed FedMaJ Algorithm}
\subsection{FedMaJ Procedure}

Algorithm~\ref{alg:fedmaj} summarizes the accuracy-adaptive FedMaJ procedure. Local training, OTA update transmission, and aggregation occur at every communication round, whereas candidate evaluation is performed only at voting rounds. Between voting rounds, the PS reuses the most recently selected rule.
\begin{algorithm}[!t]
\caption{FedMaJ with Adaptive Voting}
\label{alg:fedmaj}
\begin{algorithmic}[1]
\REQUIRE $\mathcal A$, $T$, $\mathbf w_1$, $\mathrm{Acc}_{\min}$, $\Delta_{\mathrm{acc}}$, $\tau_{\max}$
\STATE Initialize $\widetilde A_0$, $t_{\mathrm{last}}\leftarrow1$, $N_{\mathrm{vote}}\leftarrow0$; evaluate $\mathrm{Acc}_0$
\FOR{$t=1,\ldots,T$}
    \STATE Select $\mathcal S_t$, broadcast $\mathbf w_t$, and compute local updates
    \STATE Set $v_1=1$; for $t>1$, use \eqref{eq:adaptive_voting_interval}--\eqref{eq:adaptive_voting_indicator}
    \IF{$v_t=1$}
        \STATE Clients digitally upload individual updates
        \STATE PS constructs and broadcasts all $M$ candidate models using \eqref{eq:candidate_global_model}
        \STATE Clients evaluate candidates and report ordinal grades
        \STATE PS selects $\widetilde A_t=A_t^\star$ using \eqref{eq:selected_rule}
        \STATE $t_{\mathrm{last}}\leftarrow t$; $N_{\mathrm{vote}}\leftarrow N_{\mathrm{vote}}+1$
    \ELSE
        \STATE Retain $\widetilde A_t=\widetilde A_{t-1}$
    \ENDIF
    \STATE Apply $\widetilde A_t$ through \eqref{eq:ota_logical_physical_model}; update $\mathbf w_{t+1}$ and evaluate $\mathrm{Acc}_t$
\ENDFOR
\RETURN $\mathbf w_{T+1}$, rule trace, $N_{\mathrm{vote}}$
\end{algorithmic}
\end{algorithm}
At each voting round, a run uses either ordinal grades with Majority Judgment or numerical scores with average-score selection (FedMaJ-AvgScore); the two reporting schemes are never combined. The PS
also broadcasts compact control information required for normalization,
such as selected-set minima, maxima, and denominators. Each client combines
this information with its local statistics to compute
$\widetilde p_{t,k}^{(m)}$, $\widetilde\ell_{t,k}^{(m)}$,
$\widetilde c_{t,k}$, $\widetilde B_{t,k}$, and $\widetilde I_{t,k}^{-}$. Therefore, clients do not receive the raw local statistics of other participants. This limits direct information disclosure but does not provide a formal privacy guarantee.

\subsection{Communication Overhead}

At each vote, a selected client reports $M$ evaluations:
$B_{t,k}^{\mathrm{grade}}=M\lceil\log_2 n\rceil$ bits for MJ,
or $B_{t,k}^{\mathrm{score}}=Mb_{\mathrm{score}}$ bits for
AvgScore. For $S$ selected clients per round, total reporting is
$B_x^{\mathrm{ul}}=N_{\mathrm{vote}}S B_{t,k}^{x}$,
$x\in\{\mathrm{grade},\mathrm{score}\}$; with variable participation,
replace $N_{\mathrm{vote}}S$ by $\sum_t v_t|\mathcal S_t|$.
This reporting payload excludes model updates and routine metadata
(dataset sizes, participation, influence, resources, and channel
estimates), assumed available through ordinary operation.
The downlink carries one model at non-voting rounds and $M$
candidates plus $L_t^{\mathrm{ctrl}}$ control bits at voting rounds:
\begin{equation}
B_{\mathrm{total}}^{\mathrm{dl}}
=TB^{\mathrm{model}}+
\sum_{t=1}^{T}v_t\bigl[(M-1)B^{\mathrm{model}}+L_t^{\mathrm{ctrl}}\bigr].
\end{equation}
For constant control size, the additional downlink payload is
$N_{\mathrm{vote}}[(M-1)B^{\mathrm{model}}+L^{\mathrm{ctrl}}]$.

\subsection{Complexity Analysis}

For $S$ selected clients, $M$ rules, $n$ grades, and model dimension
$d_w$, local training is unchanged and ordinary aggregation costs
$O(TSd_w)$. The voting count $N_{\mathrm{vote}}=\sum_{t=1}^{T}v_t$
satisfies $\lceil T/\tau_{\max}\rceil\leq N_{\mathrm{vote}}\leq T$,
with voting rate $\rho_{\mathrm{vote}}=N_{\mathrm{vote}}/T$.
Candidate construction costs $O(N_{\mathrm{vote}}MSd_w)$ in total.
At each vote, every selected client performs $M$ model evaluations
and $O(M)$ grading operations; evaluation cost depends on the model
and local dataset. Sorting-based Majority Judgment costs
$O(N_{\mathrm{vote}}MS\log S)$, while histogram-based processing
costs $O(N_{\mathrm{vote}}M(S+n))$, linear in client--candidate
pairs for fixed $n$. Voting-schedule updates cost $O(T)$ overall.
Global-model monitoring is common to all methods and excluded
from the additional decision overhead.

\subsection{Energy Model}

We sum selected-client and PS energy over $T$ rounds and report
$\overline E^{(x)}=E^{(x)}/E_{\mathrm{ref}}$, using one fixed
$E_{\mathrm{ref}}>0$ for all components and configurations.
Physical energies are in joules; normalized values are dimensionless
model estimates, not hardware measurements. The system total is
\begin{equation}
\overline E_{\mathrm{sys}}
=\frac{1}{E_{\mathrm{ref}}}\sum_{t=1}^{T}
\left[\sum_{k\in\mathcal S_t}E_{t,k}
+E_{\mathrm{PS},t}^{\mathrm{dl}}
+E_{\mathrm{PS},t}^{\mathrm{comp}}\right].
\label{eq:normalized_total_system_energy}
\end{equation}
Client and PS computation costs are decomposed as
\begin{align}
E_{t,k}&=E_{t,k}^{\mathrm{loc}}+E_{t,k}^{\mathrm{OTA}}
+v_t\bigl(E_{t,k}^{\mathrm{eval}}+E_{t,k}^{x}\bigr),
\label{eq:client_energy_compact}\\
E_{\mathrm{PS},t}^{\mathrm{comp}}
&=E_{\mathrm{PS},t}^{\mathrm{agg}}
+v_tE_{\mathrm{PS},t}^{\mathrm{dec}},
\label{eq:ps_computation_compact}
\end{align}
where $x=\mathrm{grade}$ for FedMaJ-MJ and $x=\mathrm{score}$
for FedMaJ-AvgScore. Local training and OTA transmission occur
every round; candidate evaluation, reporting, and rule selection
occur only at voting rounds.

For any computing node $i$ and operation $o$, we use
$E_{i,t}^{o}=\kappa_i C_{i,t}^{o}f_{i,t}^{2}$, where
$\kappa_i$, $C_{i,t}^{o}$, and $f_{i,t}$ are effective switched
capacitance, CPU cycles, and frequency~\cite{yang2021energy_efficient_fl_wireless}.
This covers client training/evaluation and PS aggregation/selection.
For training, $C_{k,t}^{\mathrm{loc}}=
\nu_{t,k}n_{t,k}^{\mathrm{proc}}\chi_k^{\mathrm{cpu}}$:
local iterations times processed samples per iteration times cycles
per sample. Evaluation cycles cover all $M$ candidates.
OTA energy is $E_{t,k}^{\mathrm{OTA}}=
P_{t,k}^{\mathrm{OTA}}\tau_t^{\mathrm{OTA}}$.

For $q\in\{\mathrm{ul},\mathrm{dl}\}$, digital link rates are
\begin{equation}
r_{t,k}^{q}=W^{q}\log_2\!\left(
1+\frac{P^{q}|h_{t,k}^{q}|^2}{N_0W^{q}}\right),
\label{eq:digital_link_rate}
\end{equation}
where $W^q$, $P^q$, and $h_{t,k}^q$ denote bandwidth, transmit
power, and channel coefficient, with
$P^{\mathrm{dl}}=P_{\mathrm{PS}}^{\mathrm{dl}}$ and noise
spectral density $N_0$. Using the reporting payloads defined above,
$E_{t,k}^{x}=P^{\mathrm{ul}}B_{t,k}^{x}/r_{t,k}^{\mathrm{ul}}$.
The PS broadcasts at the weakest selected-client rate:
\begin{equation}
E_{\mathrm{PS},t}^{\mathrm{dl}}
=\frac{P_{\mathrm{PS}}^{\mathrm{dl}}
\bigl([1+v_t(M-1)]B^{\mathrm{model}}+v_tL_t^{\mathrm{ctrl}}\bigr)}
{\min_{k\in\mathcal S_t}r_{t,k}^{\mathrm{dl}}}.
\label{eq:downlink_energy_compact}
\end{equation}
Thus, fixed rules avoid voting-specific costs, while less frequent
voting reduces candidate-evaluation, reporting, control, and decision
overhead. Normalization preserves additivity and is used only for
reporting, not rule selection.
\section{Experimental Validation}
\label{sec:numerical_evaluation}

We evaluate FedMaJ using both synthetic data and the Fashion-MNIST
dataset under the same OTA-FL experimental configuration; only the
underlying data are replaced between the two experiments. The synthetic
experiment captures controlled data, resource, and wireless
heterogeneity, with proxy accuracy measured from the distance between
the global and target parameter vectors. To validate FedMaJ on real-world
data, Fig.~\ref{fig:fedmaj_MNIST} reports the
test accuracy, aggregation-influence fairness, and test loss obtained
using Fashion-MNIST. The remaining results are obtained from the
synthetic experiment. Results are averaged over ten random seeds, with
shaded regions representing one standard deviation.

\subsection{Experimental Setup}
\label{subsec:numerical_setup}

\subsubsection{Synthetic OTA-FL Environment}

We consider $K=40$ clients, select $S=7$ clients uniformly without
replacement per round, and use a $d_w=60$ dimensional model over $T=120$
rounds. Dataset sizes are sampled between $60$ and $219$, while local
targets $\mathbf w_k^{\star}$ are generated around a common center with
standard deviation $3.0$ to induce strong non-IID heterogeneity. The global
target $\mathbf w^{\star}$ is their dataset-size-weighted average. At round $t$, client $k$ computes $\Delta_{t,k}=\eta_t(\mathbf w_k^{\star}-\mathbf w_t)
+\boldsymbol\varepsilon_{t,k}$, where $\boldsymbol\varepsilon_{t,k}$ is zero-mean Gaussian noise with standard deviation $0.04$, and $\eta_t=\max\{0.10,0.22/\sqrt{1+0.015t}\}$.

The OTA channel gains follow a Gamma distribution with shape $1.0$ and scale $0.30$, clipped to $[0.02,3]$. We set $b_{t,k}\leq1.4$, $a_t=1$, and the OTA noise standard deviation to $0.02$. The communication parameters are $W^{\mathrm{ul}}=1$ MHz, $W^{\mathrm{dl}}=2$ MHz, $P^{\mathrm{ul}}=0.1$ W, $P_{\mathrm{PS}}^{\mathrm{dl}}=0.5$ W, and $N_0=10^{-12}$ W/Hz. Each model contains $B^{\mathrm{model}}=1920$ bits. The main parameters are summarized in
Table~\ref{tab:numerical_parameters}.
\begin{table*}[t]
\centering
\caption{Main parameter values.}
\label{tab:numerical_parameters}
\scriptsize
\setlength{\tabcolsep}{5.0pt}
\renewcommand{\arraystretch}{1.0}
\begin{tabular}{lclclc}
\toprule
\textbf{Parameter} & \textbf{Value} & \textbf{Parameter} & \textbf{Value} & \textbf{Parameter} & \textbf{Value} \\
\midrule
Independent seeds & $10$ &
Model dimension $d_w$ & $60$ &
Client dataset size & $60$--$219$ \\
Non-IID target standard deviation & $3.0$ &
Initial-model error standard deviation & $1.4$ &
Initial/minimum local step & $0.22/0.10$ \\
Learning-step decay coefficient & $0.015$ &
Local-update noise standard deviation & $0.04$ &
OTA-noise standard deviation & $0.02$ \\
Maximum transmit scaling & $1.4$ &
Channel Gamma shape/scale & $1.0/0.30$ &
Required accuracy $\mathrm{Acc}_{\min}$ & $0.80$ \\
Accuracy-margin step $\Delta_{\mathrm{acc}}$ & $0.01$ &
Maximum voting interval $\tau_{\max}$ & $10$ &
Number of candidates/grades $(M,n)$ & $(7,6)$ \\
Score quantization $b_{\mathrm{score}}$ & $32$ bits &
Uplink/downlink bandwidth & $1/2$ MHz &
Client/PS transmit power & $0.1/0.5$ W \\
\bottomrule
\end{tabular}\vspace{-.3cm}
\end{table*}

\subsubsection{Accuracy-Aware Grading Instantiation}

The grading-weight vector
$\boldsymbol{\alpha}_t=(\alpha_Q,\alpha_C,\alpha_R,\alpha_F)$ is adapted
to the current learning state as
\begin{equation}
\boldsymbol{\alpha}_t
=
\begin{cases}
(0.55,\,0.15,\,0.10,\,0.20),
& \mathrm{Acc}_{t-1}<\mathrm{Acc}_{\min},\\
(0.30,\,0.20,\,0.20,\,0.30),
& \text{otherwise}.
\end{cases}
\label{eq:numerical_accuracy_aware_weights}
\end{equation}
\begin{table*}[!t]
\centering
\caption{Synthetic OTA-FL proxy results. Values are mean $\pm$ standard deviation over ten seeds.}
\label{tab:fedmaj_numerical_results}
\scriptsize
\setlength{\tabcolsep}{6.0pt}
\renewcommand{\arraystretch}{1.08}
\begin{tabular}{lcccccc}
\toprule
\textbf{Method}
& \textbf{Final accuracy}
& \textbf{Normalized energy}
& \textbf{OTA distortion}
& \textbf{Fairness Jain's index}
& \textbf{$V_{\mathrm{acc}}$}
& \textbf{$N_{\mathrm{vote}}$} \\
\midrule
FedMaJ-MJ
& $0.9095\pm0.0259$
& $0.4061\pm0.0072$
& $0.0080\pm0.0007$
& $0.9535\pm0.0106$
& $0.0775\pm0.0219$
& $22.9\pm4.0$ \\
FedMaJ-AvgScore
& $0.9131\pm0.0246$
& $0.4043\pm0.0056$
& $0.0079\pm0.0006$
& $0.9524\pm0.0139$
& $0.0733\pm0.0156$
& $22.8\pm2.9$ \\
Random adaptive
& $0.9044\pm0.0391$
& $0.4523\pm0.0291$
& $0.0070\pm0.0008$
& $0.9293\pm0.0163$
& $0.1025\pm0.0518$
& $25.7\pm5.3$ \\
FedAvg
& $0.9294\pm0.0257$
& $0.4077\pm0.0034$
& $0.0081\pm0.0007$
& $0.8871\pm0.0158$
& $0.0700\pm0.0090$
& $0$ \\
Uniform aggregation
& $0.9113\pm0.0224$
& $0.3809\pm0.0028$
& $0.0074\pm0.0005$
& $0.9615\pm0.0066$
& $0.0733\pm0.0207$
& $0$ \\
Loss-aware aggregation
& $0.9107\pm0.0215$
& $0.3880\pm0.0027$
& $0.0080\pm0.0006$
& $0.9459\pm0.0130$
& $0.0700\pm0.0125$
& $0$ \\
Resource-aware aggregation
& $0.9258\pm0.0270$
& $0.3967\pm0.0034$
& $0.0074\pm0.0006$
& $0.8932\pm0.0153$
& $0.0717\pm0.0098$
& $0$ \\
Communication-aware aggregation
& $0.8621\pm0.0453$
& $0.5202\pm0.0128$
& $0.0024\pm0.0003$
& $0.8690\pm0.0143$
& $0.2450\pm0.0565$
& $0$ \\
OTA-distortion-aware aggregation
& $0.7592\pm0.0437$
& $0.9420\pm0.0422$
& $0.0013\pm0.0003$
& $0.8369\pm0.0232$
& $0.9225\pm0.0379$
& $0$ \\
Participation-aware aggregation
& $0.9286\pm0.0223$
& $0.4350\pm0.0109$
& $0.0091\pm0.0011$
& $0.9064\pm0.0112$
& $0.0658\pm0.0114$
& $0$ \\
\bottomrule
\end{tabular}\vspace{-.3cm}
\end{table*}
This prioritizes learning-quality recovery below the accuracy threshold and restores a balanced learning--communication--resource--fairness tradeoff afterward.

\noindent To prevent a candidate with insufficient learning quality from winning only because of favorable resource or fairness indicators, Majority Judgment is applied to the learning-safe set
\begin{equation}
    \mathcal A_t^{\mathrm{safe}}
    =
    \left\{
    A_m\in\mathcal A:
    \underset{k\in\mathcal S_t}{\operatorname{median}}\,
    Q_{t,k}^{(m)}\geq Q_{t,k,m}^{\mathrm{safe}}=0.55
    \right\}.
    \label{eq:numerical_safe_candidate_set}
\end{equation}
If this set is empty, the candidate with the smallest mean client-local
candidate loss is retained as an emergency fallback. The six ordinal grades
are Very Poor, Poor, Fair, Good, Very Good, and Excellent, with score
thresholds
$\{0,0.40,0.55,0.65,0.75,0.85,1\}$.
\subsubsection{Compared Methods}

We compare FedMaJ-MJ with FedMaJ-AvgScore, which uses the same candidate library, indicators, safety filter, and adaptive voting schedule but averages the numerical client scores, and Random adaptive, which randomly selects a rule at each adaptive voting round. We also consider fixed rules $A_1$--$A_7$, each applied over the entire training horizon. All methods use identical selected-client sequences, channels, initial models, update noise, and OTA noise for each seed. For readability, the figures show representative fixed rules, while Table~\ref{tab:fedmaj_numerical_results} reports all fixed-rule results.

\subsection{Evaluation Metrics}
\label{subsec:numerical_metrics}

The synthetic test loss and proxy accuracy are
\begin{equation}
L_t^{\mathrm{test}}
=
\frac{1}{d_w}
\left\|\mathbf w_{t+1}-\mathbf w^{\star}\right\|_2^2,
\qquad
\mathrm{Acc}_t
=
\frac{1}{1+L_t^{\mathrm{test}}}.
\label{eq:proxy_accuracy}
\end{equation}
The OTA distortion is
$D_t=d_w^{-1}\|\Delta_t-\widehat{\Delta}_t\|_2^2$.
Given the cumulative client influence $I_{t,k}$, Jain fairness writes
\begin{equation}
J_t^{\mathrm{inf}}
=
\frac{\left(\sum_{k=1}^{K}I_{t,k}\right)^2}
{K\sum_{k=1}^{K}I_{t,k}^2}.
\label{eq:numerical_jain_fairness}
\end{equation}
The accuracy-threshold violation rate is estimated by
\begin{equation}
    V_{\mathrm{acc}}
    =
    \frac{1}{T}
    \sum_{t=1}^{T}
    \mathbb{I}\!\left[
    \mathrm{Acc}_t<\mathrm{Acc}_{\min}
    \right].
    \label{eq:numerical_violation_rate}
\end{equation}

At voting rounds, we report client satisfaction
\eqref{eq:client_satisfaction}, its client--vote mean
\eqref{eq:overall_client_satisfaction}, and the top-choice rate
\eqref{eq:client_top_choice_rate}. These diagnostics apply to FedMaJ-MJ,
FedMaJ-AvgScore, and random adaptive selection because they evaluate all
candidates. For fixed rules, candidate grades are reconstructed offline
to compute \eqref{eq:overall_active_rule_client_satisfaction}, without
adding modeled overhead or energy.

Energy is normalized using the common reference defined in the energy
model, chosen as the maximum final physical system energy across all
methods, seeds, and fixed-interval configurations.

\begin{figure*}[!t]
    \centering
    \setlength{\tabcolsep}{1pt}
    \begin{tabular}{ccc}
        \includegraphics[width=0.33\textwidth]
        {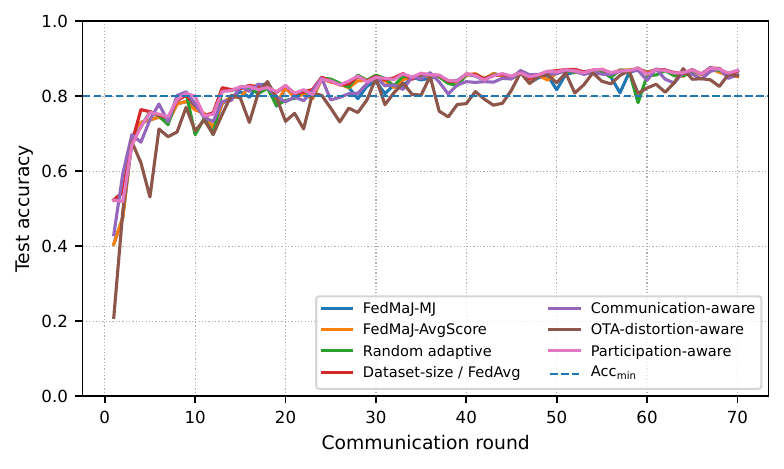}
        &
        \includegraphics[width=0.33\textwidth]
        {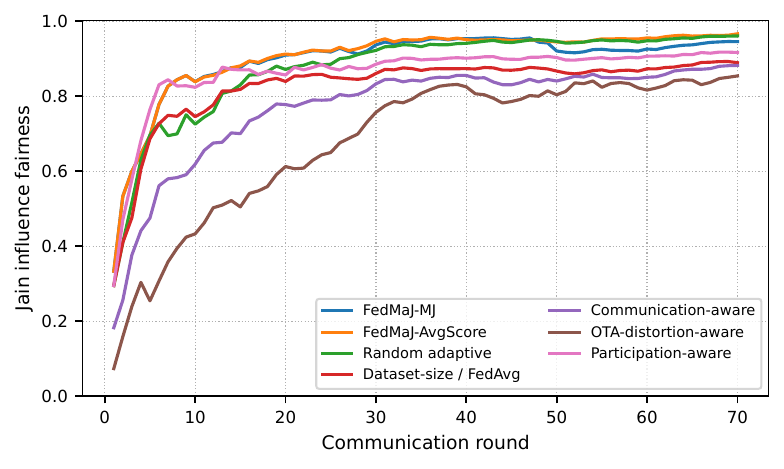}
        &
        \includegraphics[width=0.33\textwidth]
        {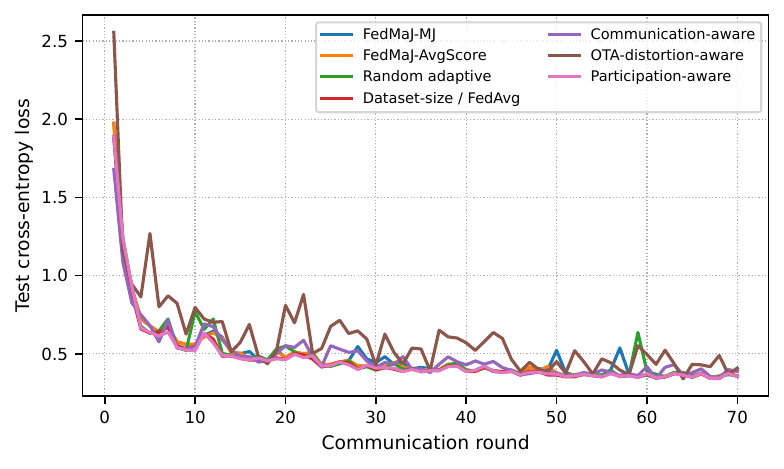}
        \\
        {\footnotesize (a) Test accuracy }
        &
        {\footnotesize (b) Jain aggregation-influence fairness}
        &
        {\footnotesize (c) Test loss }
    \end{tabular}
    \caption{Fashion-MNIST evaluation of FedMaJ and the considered baselines
in terms of test accuracy, Jain aggregation-influence fairness, and test
loss over communication rounds.}
    \label{fig:fedmaj_MNIST}
    \vspace{-.3cm}
\end{figure*}

\subsection{Results and Discussion}
\label{subsec:numerical_results}

\subsubsection{Synthetic Learning and Fairness}
\begin{figure}[!t]
    \centering
    \begin{subfigure}[t]{0.48\columnwidth}
        \centering
        \includegraphics[width=\linewidth]
        {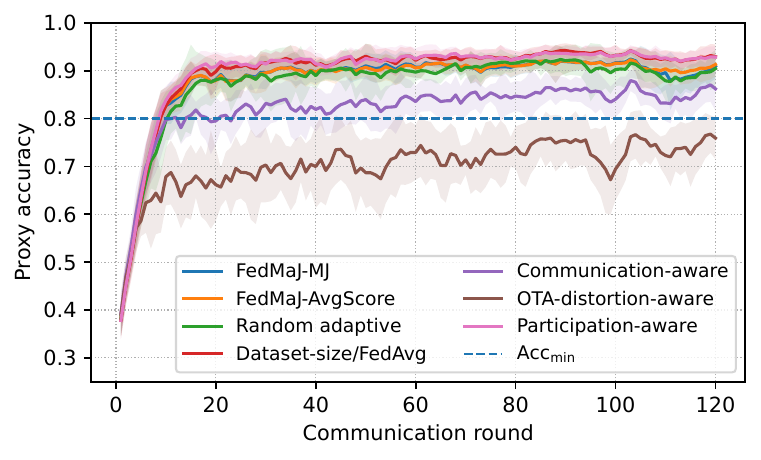}
        \caption{Proxy accuracy.}
        \label{fig:fedmaj_accuracy}
    \end{subfigure}
    \hfill
    \begin{subfigure}[t]{0.48\columnwidth}
        \centering
        \includegraphics[width=\linewidth]
        {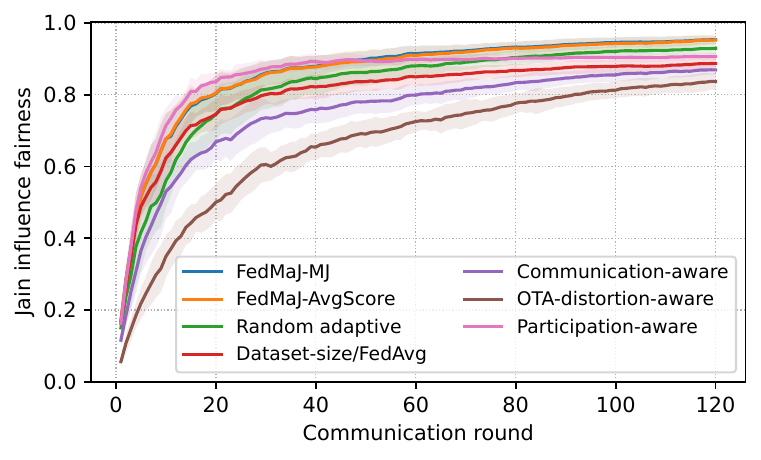}
        \caption{Influence fairness.}
        \label{fig:fedmaj_fairness}
    \end{subfigure}
    \caption{Synthetic OTA-FL proxy accuracy and Jain influence
    fairness over ten seeds (mean $\pm$ standard deviation).}
    \label{fig:synthetic_learning_fairness}
\end{figure}

Most learning-oriented methods reach the accuracy threshold within
$10$--$20$ rounds (Fig.~\ref{fig:synthetic_learning_fairness}(a)).
FedMaJ-MJ achieves $0.9095\pm0.0259$, close to FedMaJ-AvgScore
($0.9131\pm0.0246$), exceeding communication-aware and
OTA-distortion-aware aggregation by $4.74$ and $15.03$ percentage
points. Its influence fairness reaches $0.9535\pm0.0106$, close
to uniform aggregation and $7.48\%$ above FedAvg, versus $0.9293$
for random selection (Fig.~\ref{fig:synthetic_learning_fairness}(b)).
Violation rates are $7.75\%$, $7.33\%$, and $10.25\%$ for
FedMaJ-MJ, FedMaJ-AvgScore, and random selection, versus $24.50\%$
and $92.25\%$ for the two wireless-oriented rules
(Fig.~\ref{fig:fedmaj_additional_evaluation}(b)). FedAvg and
participation-aware aggregation achieve slightly fewer violations,
but lower influence fairness than FedMaJ-MJ.

\subsubsection{Client Preferences and Robustness}
\begin{figure}[!t]
    \centering
    \setlength{\tabcolsep}{1pt}
    \begin{tabular}{cc}
        \includegraphics[width=0.49\columnwidth]
        {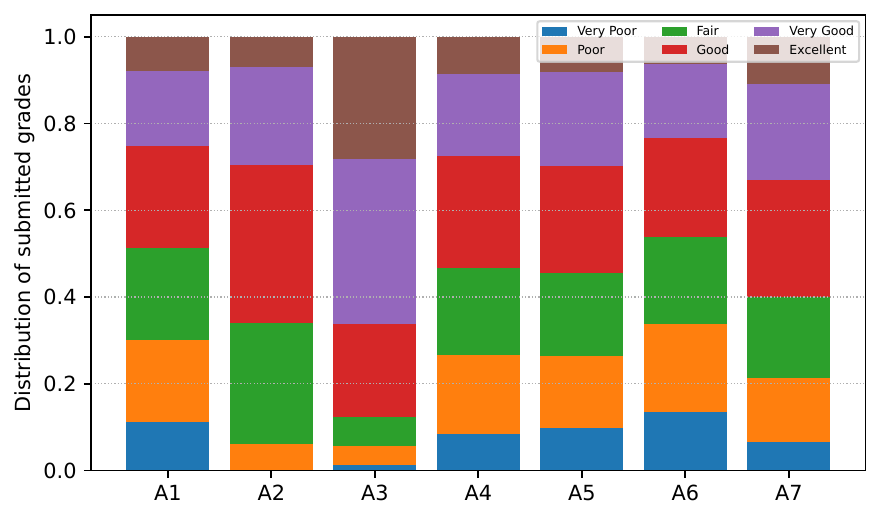}
        &
        \includegraphics[width=0.49\columnwidth]
        {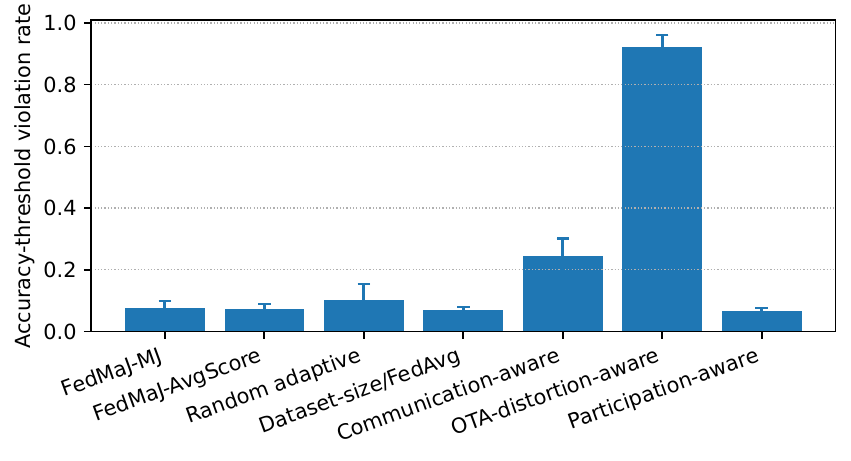}
        \\
        {\footnotesize (a) Ordinal-grade distributions}
        &
        {\footnotesize (b) Threshold violation rate}
        \\[0.1cm]
        \includegraphics[width=0.49\columnwidth]
        {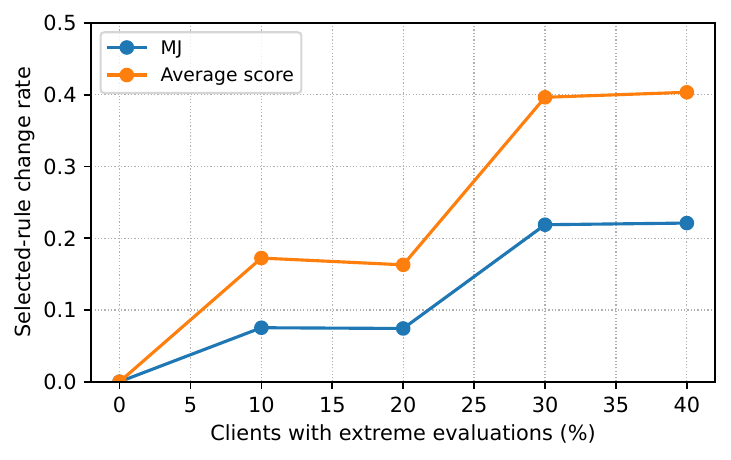}
        &
        \includegraphics[width=0.49\columnwidth]
        {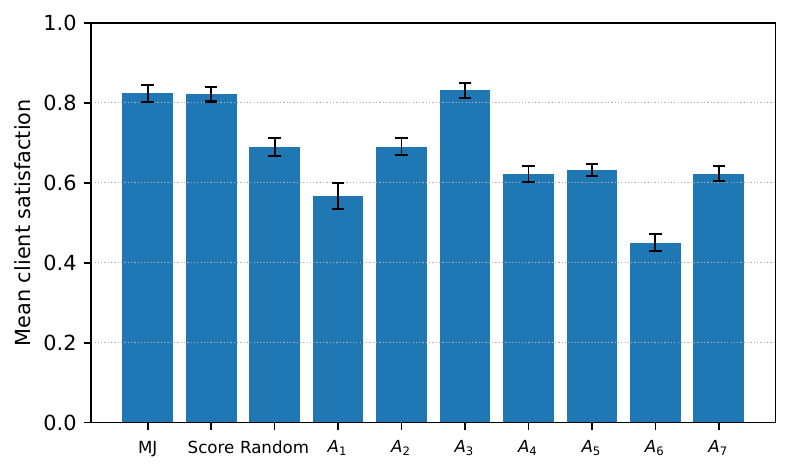}
        \\
        {\footnotesize (c) Robustness to extreme evaluations}
        &
        {\footnotesize (d) Client satisfaction}
    \end{tabular}
    \caption{Evaluation of FedMaJ in terms of ordinal-grade distributions, accuracy-threshold violations, robustness to extreme
    client evaluations, and active-rule client satisfaction.}
    \label{fig:fedmaj_additional_evaluation}
    \vspace{-.3cm}
\end{figure}

Loss-aware aggregation receives the most Very Good and Excellent
grades, while OTA-distortion-aware aggregation receives the most
Very Poor and Poor grades (Fig.~\ref{fig:fedmaj_additional_evaluation}(a)).
These distributions pool rounds and seeds; voting uses round-specific
profiles and MJ tie-breaking. FedMaJ-MJ, FedMaJ-AvgScore, and fixed
$A_3$ show similar offline satisfaction, consistent with $A_3$
dominating selections (Fig.~\ref{fig:fedmaj_additional_evaluation}(d)).
We test sensitivity by assigning maximum scores to one non-winning
candidate for a fraction of clients at saved votes. At $10\%$
perturbation, MJ reduces the rule-change rate from $0.1723$ to
$0.0753$ ($56.3\%$); at $30\%$, rates fall from $0.3964$ to $0.2188$
(Fig.~\ref{fig:fedmaj_additional_evaluation}(c)), indicating lower
sensitivity to the tested extreme evaluations.

\subsubsection{Fashion-MNIST Validation}
Fig.~\ref{fig:fedmaj_MNIST} shows that FedMaJ-MJ and FedMaJ-AvgScore
converge to accuracy and loss levels comparable to learning-oriented
fixed rules while maintaining high influence fairness. FedMaJ-MJ
satisfies the accuracy threshold and distributes cumulative influence
more evenly than several fixed rules. Random adaptive selection is
more variable, while wireless-oriented rules retain higher losses.
Together, these results support adaptive rule selection as a way to
balance learning and influence fairness, with MJ additionally reducing
sensitivity to extreme client evaluations.
\section{Conclusion}

This paper introduced FedMaJ, which dynamically selects aggregation rules via client grading and Majority Judgment, with an accuracy-adaptive voting schedule. Synthetic OTA-FL evaluations demonstrate competitive proxy accuracy and superior influence fairness compared to FedAvg. Fashion-MNIST experiments further confirm stable learning performance and high fairness. Under extreme-evaluation perturbations, Majority Judgment reduces the rule-change rate by $56.3\%$ over average-score selection, establishing ordinal voting as a robust control layer. These findings highlight ordinal consensus as an effective mechanism for balancing client preferences and system objectives. Future work will explore larger datasets, complex models, time-varying channels, and strategic grading.

\bibliographystyle{IEEEtran}
\bibliography{references}

@article{amiri2020federated,
  title        = {Federated Learning Over Wireless Fading Channels},
  author       = {Amiri, Mohammad Mohammadi and G{\"u}nd{\"u}z, Deniz},
  journal      = {IEEE Transactions on Wireless Communications},
  volume       = {19},
  number       = {5},
  pages        = {3546--3557},
  year         = {2020},
  publisher    = {IEEE}
}

@inproceedings{McMahan2017,
  author    = {Brendan McMahan and Eider Moore and Daniel Ramage and Seth Hampson and Blaise Ag{\"u}era y Arcas},
  title     = {Communication-Efficient Learning of Deep Networks from Decentralized Data},
  booktitle = {Proceedings of the 20th International Conference on Artificial Intelligence and Statistics},
  series    = {Proceedings of Machine Learning Research},
  volume    = {54},
  pages     = {1273--1282},
  year      = {2017},
  publisher = {PMLR}
}

@INPROCEEDINGS{zhang2025dual,
  author={Zhang, Haizhou and Yu, Xianjia and Westerlund, Tomi},
  booktitle={IEEE 11th World Forum on Internet of Things (WF-IoT)}, 
  title={Dual-Criterion Model Aggregation in Federated Learning: Balancing Data Quantity and Quality}, 
  year={2025},
  volume={},
  number={},
  pages={1-3}
  }

@inproceedings{nishio2018client,
  title={Client Selection for Federated Learning with Heterogeneous Resources in Mobile Edge},
  author={Nishio, Takayuki and Yonetani, Ryo},
  booktitle={IEEE International Conference on Communications},
  year={2019}
}

@article{tan2022adafed,
  title={AdaFed: Optimizing Participation-Aware Federated Learning With Adaptive Aggregation Weights},
  author={Tan, Lei and Zhang, Xiaoxi and Zhou, Yipeng and Che, Xinkai and Hu, Miao and Chen, Xu and Wu, Di},
  journal={IEEE Transactions on Network Science and Engineering},
  volume={9},
  number={4},
  pages={2708--2720},
  year={2022},
  publisher={IEEE},
  doi={10.1109/TNSE.2022.3168969}
}

@article{shen2025fedmcon,
  title = {FedMcon: an adaptive aggregation method for federated learning via meta controller},
  volume = {26},
  number = {8},
  journal = {Frontiers of Information Technology \&amp; Electronic Engineering},
  publisher = {Zhejiang University Press},
  author = {Shen,  Tao and Li,  Zexi and Zhao,  Ziyu and Zhu,  Didi and Lv,  Zheqi and Kuang,  Kun and Zhang,  Shengyu and Wu,  Chao and Wu,  Fei},
  year = {2025},
  month = Aug,
  pages = {1378–1393}
}

@article{yang2020ota_fl,
  title = {Federated Learning via Over-the-Air Computation},
  volume = {19},
  number = {3},
  journal = {IEEE Transactions on Wireless Communications},
  publisher = {Institute of Electrical and Electronics Engineers (IEEE)},
  author = {Yang,  Kai and Jiang,  Tao and Shi,  Yuanming and Ding,  Zhi},
  year = {2020},
  month = Mar,
  pages = {2022–2035}
}

@article{cao2022air_fedavg_power_control,
  author={Cao, Xiaowen and Zhu, Guangxu and Xu, Jie and Cui, Shuguang},
  journal={IEEE Journal on Selected Areas in Communications}, 
  title={Transmission Power Control for Over-the-Air Federated Averaging at Network Edge}, 
  year={2022},
  volume={40},
  number={5},
  pages={1571-1586},
  doi={10.1109/JSAC.2022.3143217}
}

@article{sun2022dynamic_scheduling_airfeel,
  author = {Sun, Yuyi and Zhou, Sheng and Niu, Zhisheng and G{\"u}nd{\"u}z, Deniz},
  title = {Dynamic Scheduling for Over-the-Air Federated Edge Learning with Energy Constraints},
  journal = {IEEE Journal on Selected Areas in Communications},
  volume = {40},
  number = {1},
  pages = {227--242},
  year = {2022},
  doi = {10.1109/JSAC.2021.3126056}
}

@article{azimi2024wafel,
  author={Azimi-Abarghouyi, Seyed Mohammad and Tassiulas, Leandros},
  journal={IEEE Transactions on Wireless Communications}, 
  title={Over-the-Air Federated Learning via Weighted Aggregation}, 
  year={2024},
  volume={23},
  number={12},
  pages={18240-18253},
  doi={10.1109/TWC.2024.3463754}
  }

@article{hamidi2025ota_ffl,
  author={Mohajer Hamidi, Shayan and Bereyhi, Ali and Asaad, Saba and Vincent Poor, H.},
  journal={IEEE Communications Letters}, 
  title={Over-the-Air Fair Federated Learning via Multi-Objective Optimization}, 
  year={2025},
  volume={29},
  number={7},
  pages={1549-1553},
  doi={10.1109/LCOMM.2025.3567387}
  }

@article{yang2021energy_efficient_fl_wireless,
  author  = {Yang, Zhaohui and Chen, Mingzhe and Saad, Walid and Hong, Choong Seon and Shikh-Bahaei, Mohammad},
  title   = {Energy Efficient Federated Learning Over Wireless Communication Networks},
  journal = {IEEE Transactions on Wireless Communications},
  volume  = {20},
  number  = {3},
  pages   = {1935--1949},
  year    = {2021},
  doi     = {10.1109/TWC.2020.3037554}
}

@book{MJbook,
    author = {Balinski, Michel and Laraki, Rida},
    title = {Majority Judgment: Measuring, Ranking, and Electing},
    publisher = {The MIT Press},
    year = {2011},
    month = {01}
}
\end{document}